\documentclass[12pt]{iopart}

\usepackage{graphicx}        
\usepackage{amssymb}        
\renewcommand{\vec}[1]{\mathbf{\mathit #1}}

\begin{document}


\title[]{Large-scale Model of the Axisymmetric Dynamo with Feedback Effects}

\author{Laura Sraibman$^{1,2}$ and Fernando Minotti$^{1,2}$}
             
\address{$^1$Universidad de Buenos Aires, Facultad de Ciencias Exactas y
Naturales, Departamento de F\'{\i}sica, Buenos Aires, Argentina.\\$^2$CONICET-Universidad de Buenos Aires, Instituto de F\'{\i}sica del Plasma (INFIP), Buenos Aires, Argentina}

\begin{abstract}
A dynamo model is presented, based on a previously introduced kinematic model, in which the reaction of the magnetic field on the mass flow through the Lorentz force is included. Given the base mass flow corresponding to the case with no magnetic field, and assuming that the modification of this flow due to the Lorentz force can be treated as a perturbation, a complete model of the large-scale magnetic field dynamics can be obtained. The input needed consists in the large-scale meridional and zonal flows, the small-scale magnetic diffusivity, and a constant parameter entering the expression of the $\alpha$-effect. When applied to a solar-like star, the model shows a realistic dynamics of the magnetic field, including cycle duration, consistent field amplitudes with the correct parity, progression of the zonal magnetic field towards the equator, and motion toward the poles of the radial field at high latitudes. Also, the radial and zonal components show a correct phase relation, and, at the surface level, the magnetic helicity is predominantly negative in the northern hemisphere and positive in the southern hemisphere.  
\end{abstract}

\ead{lsraibman@df.uba.ar}
\ead{minotti@df.uba.ar}
\maketitle

\section{Introduction}
     \label{S-Introduction} 

Mean field dynamo models are valuable tools to test and understand the
physical mechanisms involved in the dynamics of the mean magnetic field in
astrophysical objects. Kinematic dynamo models with parametrized $\alpha$ and/or Babcock-Leighton mechanisms \cite{Karak2016,Belucz2015} are useful to test prescribed matter flows in its ability to sustain or amplify a magnetic field. In particular, flux-transport dynamos, $\alpha-\Omega$ models coupled with meridional mass flows, have been very successful in explaining many of the observed features of the large scale magnetic field in the Sun \cite{Wang1991, Choudhuri1995, Durney1995, Dikpati1999, Kuker2001, Bonanno2002, Guererro2004, Jouve2007}. Also, when properly complemented with models of feedback effects of the magnetic field, and eventually some addition of stochasticity, flux-transport dynamos are also able to more fully reproduce realistic dynamics \cite{Rempel2006b,Passos2014, Hazra2017, Lemerle2017}. More fundamental approaches, like the use of the full set of 3-D magnetohydrodynamic (MHD) large-scale equations in the convection zone of the star \cite{Miesch2014,Passos2015,Miesch2016}, or 2-D azimuthally averaged versions \cite{Kitchatinov2017}, solve the coupled dynamics of magnetic field and mass flow, thus allowing to determine from more fundamental principles the large scale flows and the magnetic field interrelation. These latter approaches still require modeling of the effect of turbulent small scales on the large scales reproduced by the numerical simulation. A validation of the modeling used in either approach is possible through comparison of the results of the theory with observations, particularly of the Sun for which the most accurate and extensive observations are possible. We consider here an approach intermediate between mean-field kinematic dynamos and full 3-D MHD simulations, consisting in a 2-D axisymmetric dynamo model with prescribed base flows, corresponding to zero magnetic field, coupled with dynamic equations describing the back reaction of the magnetic field on the mass flow. The axisymmetric large-scale dynamo model is that presented in a recent work, in which the $\alpha$ and diffusivity tensors are determined by the large-scale flows, using a novel technique for deriving large-scale equations directly from the original equations, employing only first principles \cite{Minotti2000}. This technique allows us to separate the problem into three scales: large scales resolved by the numerical procedure, intermediate scales whose dynamic effect on the resolved scales is modeled by the mentioned technique, and microscales that are described separately by simpler models.  We here show that when the reaction of the magnetic field on the meridional flow through the Lorentz force is considered, the consequent modifications of the $\alpha$ and diffusivity tensors is sufficient to ensure a realistic dynamics of the magnetic field when applied to a solar-like star, including cycle period, field amplitudes and phases, correct shift towards equator and poles of the zonal magnetic field at mid-latitudes, and of the radial field at high latitudes, respectively. Also, the predominant magnetic helicity has the correct sign in each hemisphere at the surface level \cite{Charbonneau2010}, which are features observed at the surface that constrain the internal dynamics of the magnetic field \cite{Cameron2015}.
 
\section{Basic Formalism}
      \label{S-basics}      
We give here a very brief presentation of the technique in \cite{Minotti2000} as applied in \cite{Sraibman} for the derivation of equations for the large-scale fields from the original dynamic equations. The large scale component of a generic field $c(\mathbf{\mathit{x}},t)$ is determined by the running volume average 
\begin{equation}
C(\vec{X},t)=\left\langle c(\vec{x},t)\right\rangle _{\vec{X}}=
\frac{1}{\triangle V}\int c(\vec{x},t)\textrm{d}V,  \label{average}
\end{equation}
in which $\vec{X=}\left\langle \vec{x}\right\rangle _{\vec{X}}$
denotes the center of the volume $\triangle V$, whose linear length $
\lambda $, defines the size of the large scale. 
Fluctuations of $c(\vec{x},t)$ are defined in non-standard form as
\begin{equation}
\delta c(\vec{X},\vec{x},t)=c(\vec{x},t)-C(\vec{X},t),
\label{fluct}
\end{equation}
based on Schumann's prescription \cite{Schumann} to avoid the appearance of Leonard's stresses and cross-terms \cite{Leonard,Clark} in the final equations. With definitions \ref{average} and \ref{fluct} the averages satisfy Reynolds' postulates, 
\begin{equation}
\left\langle C(\vec{X})\right\rangle _{\vec{X}}=C(\vec{X}
),\;\;\;\left\langle \delta c(\vec{X},\vec{x},t)C(\vec{X})\right\rangle
_{\vec{X}}=0,  \label{propert}
\end{equation}
and 
\begin{equation}
\left\langle \frac{\partial c}{\partial \vec{x}}\right\rangle _{\vec{X}
}=\frac{\partial C}{\partial \vec{X}}.  \label{deri}
\end{equation}
These relations allow to very simply apply the averaging procedure to the  equation of magnetic induction: 
\begin{equation}
\frac{\partial \vec{b}}{\partial t}=\nabla \times \left( \vec{u}\times 
\vec{b}-\eta \nabla \times \vec{b}\right) ,  \label{beqs}
\end{equation}
in which $\eta $ denotes the microscale turbulent diffusivity, while $\vec{b}$ and $\vec{u}$ represent the magnetic and velocity fields, respectively. The result is 
\begin{equation}
\frac{\partial \vec{B}}{\partial t}=\nabla \times \left( \vec{U}\times 
\vec{B}-\eta \nabla \times \vec{B}\right) +\nabla \times \vec{S},\
\label{beqav}
\end{equation}
with capital letters denoting large-scale components of the fields represented by the corresponding lower-case letters. The term $\vec{S}$ corresponds to the effect of scales smaller than  $\lambda $ on the large-scale field, written as 
\begin{equation}
\vec{S}(\vec{X})=\left\langle \delta \vec{u}(\vec{X},\vec{x}
)\times \delta \vec{b}(\vec{X},\vec{x})\right\rangle _{\vec{X}}.
\label{tau}
\end{equation}
The vector $\vec{S}$, expressed in terms of the large-scale fields, is obtained as described in detail in \cite{Minotti2000}. Following that work it is determined that for generic fields $a({\vec{x}})$ and $c({\vec{x}})$, with averages $A({\vec{X}})$ and $C({\vec{X}})$, and fluctuations $\delta a({\vec{X}},{\vec{x}})$ and $\delta c({\vec{X}},{\vec{x}})$, as defined above, one has in general at leading order in the parameter $\lambda/L$, where $L$ is the characteristic length scale of the large-scale magnitudes,
\begin{equation}
\left\langle \delta a({\vec{X}},{\vec{x}})\delta c({\vec{X}},{\vec{x}})\right\rangle
_{{\vec{X}}} =\frac{\lambda ^{2}}{24}\nabla _{{\vec{X}}}A\cdot \nabla _{{\vec{X}}
}C.  \label{general}
\end{equation}
In this way, the vector $\vec{S}$ is expressed in Cartesian components as 
\begin{equation}
S_{i}(\vec{X})=\frac{\lambda ^{2}}{24}\varepsilon _{ijk}\frac{\partial
U_{j}}{X_{m}}\frac{\partial B_{k}}{X_{m}},  \label{a}
\end{equation}
where $\varepsilon _{ijk}$ is Levi-Civita's pseudo-tensor.

The general expression \ref{general} is derived from exact relations between large-scale magnitudes corresponding to two different scales, with the only assumption that the large-scale fields are spatially smooth on scale lengths of order $\lambda$. This condition can be verified \textit{a posteriori}, and in order to satisfy it the spatial grid size for the numerical simulation is taken very close to $\lambda$. In this respect, the approach followed here is not that of modeling of turbulence, but rather that of subgrid modeling. The advantage of the approach is that relation \ref{general} captures particularly well the effect of scales not resolved by the numerical simulation, intermediate between resolved scales and small ones. The resulting separation of scales is helpful in order to effectively model the dynamic effects of small scale turbulence on resolved scales \cite{Rudiger2004}.          

This is the basis of the kinematic dynamo model presented in \cite{Sraibman}, which includes a further average on the azimuthal angle $\phi$ around the rotation axis $z$ in order to reduce it to an axisymmetric model. Using spherical coordinates $r, \theta , \phi $, and time $t$, the large-scale magnetic field $\vec{B}$\ so averaged is represented using its azimuthal component $B_{\phi}(r,\theta ,t)$ and the zonal vector potential $A_{\phi}(r,\theta ,t)$, 
\begin{equation}
\vec{B}=\nabla \times \left( A_{\phi}\,\textbf{e}_{\phi }\right) +B_{\phi}\,\textbf{e}_{\phi }.  \label{Bavg}
\end{equation}
The evolution equations are finally written as 
\begin{equation}
\frac{\partial A_{\phi}}{\partial t}=U_{r}B_{\theta }-U_{\theta }B_{r}+\eta \left(
\nabla ^{2}A_{\phi}-\frac{A_{\phi}}{r^{2}\sin ^{2}\theta }\right) +S_{\phi }+\alpha B_{\phi},
\label{dAdt}
\end{equation}
and 
\begin{equation}
\frac{\partial B_{\phi}}{\partial t}=\left[ \nabla \times \left( \vec{U}\times 
\vec{B}-\eta \nabla \times \vec{B}\right) +\nabla \times \vec{S}
\right] \cdot \textbf{e}_{\phi },  \label{dBdt}
\end{equation}
with prescribed (microscale) magnetic diffusivity $\eta$ and
axisymmetric large-scale velocity $\vec{U}$. The latter includes a
meridional velocity $\vec{U}_{\mathrm{m}}$, and an azimuthal component $U_{\phi
}=$ $\Omega \left( r,\theta \right) r\sin \theta $, where $\Omega \left(
r,\theta \right) $ is the local large-scale angular velocity. The vector $\vec{S}$, given in terms of explicit expressions of $\vec{U}$ and $\vec{B}$, corresponds to the diffusivity and $\alpha $ tensor
components other than $\alpha _{\phi \phi }$, which is explicitly
written as the last term in Equation \ref{dAdt}, with $\alpha _{\phi \phi
}\equiv \alpha $.

From the derivations in \cite{Sraibman} it is also shown that the $\alpha$-coefficient can be expressed in terms of the radial cylindrical component of the mean vorticity, $\omega _{s}$, 
\begin{equation}
\alpha =\frac{\lambda ^{2}\varkappa }{24s}\omega _{s},  \label{alfa}
\end{equation}
where $0<\varkappa <1$ is a parameter of the model, and $s=r\sin \theta $ is the radial cylindrical coordinate.

The more involved expressions of the terms dependent on the vector $\vec{S}$, as derived in \cite{Sraibman}, are presented in the Appendix. 
    
\section{Inclusion of Reaction Effects} 
      \label{S-reaction}      
Note that $\vec{U}$ not only appears in the explicit advection terms in Equations \ref{dAdt} and \ref{dBdt}, but also enters in the determination of $\vec{S}$ and  $\alpha $. In this way, as the magnetic field modifies the flow through the action of the Lorentz force, its effect on the magnetic field evolution represents a possible self-regulating mechanism.

To account for this effect in the simplest possible way, we consider that the modification of the flow due to the Lorentz force can be
taken as a perturbation to the imposed base flow, a condition that can
be tested \textit{a posteriori}, and which is well satisfied in the application to a solar-like star presented here. A very similar approach in this respect is used in \cite{Hazra2017} to study the variation of the meridional circulation in the Sun due the changing Lorentz force during the solar cycle. We also note that feedback effects of the magnetic field in dynamo models are being studied since a long time \cite{Brandenburg1990, Brandenburg1992, Kuker1999, Rempel2006b}. These works include magnetic feedback effects on the differential rotation and/or the meridional flow, that directly reflects on the $\Omega$-effect and on the convection of magnetic lines by the mean flow. Feedback effects on the  $\alpha$ effect are usually modeled as an $\alpha$-quenching for large enough magnitudes of the toroidal field \cite{Rempel2006b}. The main difference in the present approach is the model employed that directly relates the $\alpha$ and diffusivity tensors to the large-scale flow and their consequent modification by the Lorentz force.     

To evaluate the effect of the Lorentz force on the flow we consider the
stationary version of the $z$-component of the angular momentum evolution
equation 
\begin{eqnarray}
\rho \vec{U}_{\mathrm{m}}\cdot \nabla \left[ r^{2}\sin ^{2}\theta \Omega \left(
r,\theta \right) \right] +\nabla \cdot \left[ r\sin \theta \left\langle \delta u_{\phi }\delta
\left( \rho \vec{u}_{\mathrm{m}}\right) \right\rangle \right] \nonumber  \\
=\frac{1}{\mu _{0}}\left[ \nabla \times \left( A_{\phi}\,\textbf{e}_{\phi
}\right) \right] \cdot \nabla \left( r\sin \theta B_{\phi}\right) ,\label{Lz}
\end{eqnarray}
where $\mu _{0}$ is the vacuum permeability, and $\delta u_{\phi }$ and $
\delta \left( \rho \vec{u}_{\mathrm{m}}\right) $ correspond to the fluctuations of
the azimuthal velocity and meridional mass flow, respectively, as defined in Equation \ref{fluct}. The right-hand side of Equation \ref{Lz} corresponds to the $z$ component of the large scale Lorentz force torque. When this torque is absent the balance is satisfied by base values of flows and fluctuations, indicated by an upper index (0), 
\begin{equation}
\rho \vec{U}_{\mathrm{m}}^{(0)}\cdot \nabla \left[ r^{2}\sin ^{2}\theta \Omega
^{(0)}\left( r,\theta \right) \right]+\nabla \cdot \left[ r\sin \theta \left\langle \delta u_{\phi }\delta
\left( \rho \vec{u}_{\mathrm{m}}\right) \right\rangle ^{(0)}\right]=0.
\end{equation}

It is then assumed that the magnetic torque generates a perturbation to the
meridional flow, $\vec{U}_{\mathrm{m}}^{(1)}$, required to establish the balance
in Equation \ref{Lz}. Moreover, with the formalism developed in \cite{Minotti2000}, similarly to Equation \ref{a}, we can write the average of these fluctuations as 
\begin{equation}
\left\langle \delta u_{\phi }\delta \left( \rho \vec{u}_{\mathrm{m}}\right)
\right\rangle ^{(1)}\simeq \frac{{\lambda }^{2}}{24}\nabla (\Omega _{0}r\sin
\theta )\cdot \nabla \left( \rho \vec{U}_{\mathrm{m}}^{(1)}\right) .\nonumber
\end{equation}
The order of magnitude of this average can be estimated considering that for a generic large-scale magnitude $F$ one has $\lambda \left\vert \nabla F\right\vert \approx F$ so that, due to the factor $1/24$, the average is about an order of magnitude smaller than the first term in the left-hand side of Equation \ref{Lz}, and can thus be neglected. In this way we have
\begin{eqnarray}
\rho \vec{U}_{\mathrm{m}}^{(1)}\cdot \nabla \left[ r^{2}\sin ^{2}\theta \Omega
^{(0)}\left( r,\theta \right) \right]
+\rho \vec{U}_{\mathrm{m}}^{(0)}\cdot \nabla \left[ r^{2}\sin ^{2}\theta \Omega
^{(1)}\left( r,\theta \right) \right] \nonumber \\
=\frac{1}{\mu _{0}}\left[ \nabla \times \left( A_{\phi}\,\textbf{e}_{\phi
}\right) \right] \cdot \nabla \left( r\sin \theta B_{\phi}\right). \label{um1intermediate}
\end{eqnarray}
This equation is of course not sufficient to determine both perturbed
fields, $\vec{U}_{\mathrm{m}}^{(1)}$ and $\Omega ^{(1)}$. To proceed we note that the order of the second term in the left-hand side of \ref{um1intermediate} is related to the order of the first one by a factor $f$:
\begin{equation}
f=\frac{\left\vert\vec{U}_{\mathrm{m}}^{(0)}\Omega^{(1)}\right\vert}
{\left\vert\vec{U}_{\mathrm{m}}^{(1)}\Omega_{0}\right\vert}.
\end{equation}
In the particular case of the Sun, the observed relative variation of the meridional flow during the solar cycle is about 25\% \cite{Hathaway2010}, while that of the angular velocity is about 1\% \cite{Howe2000,Thompson2004}, leading to $f \approx 0.04$.     
Relating these variations to the effect of the Lorentz force one can also obtain an priori estimation taking into account that the meridional components of the Lorentz force have terms of order $B_{\phi}^{2}$ and of order $\left\vert \nabla \times \left( A_{\phi}\,\textbf{e}_{\phi}\right) \right\vert ^{2}$, whereas the azimuthal component is of order $\left\vert \nabla \times \left( A_{\phi}\,\textbf{e}_{\phi }\right)\right\vert \left\vert B_{\phi}\right\vert $. We thus see that if either of $\left\vert B_{\phi}\right\vert $ or $\left\vert \nabla \times \left( A_{\phi}\,\textbf{e}_{\phi }\right) \right\vert $ prevails over the other, the meridional flow is expected to be more perturbed than the azimuthal flow, whereas if $\left\vert B_{\phi}\right\vert \approx \left\vert \nabla \times \left( A_{\phi}\,\textbf{e}_{\phi }\right) \right\vert $, both components of the flow are expected to be equally affected. In this way one should correspondingly have either $\left\vert \vec{U}_{\mathrm{m}}^{(1)}\right\vert \gg \left\vert \Omega ^{(1)}\right\vert R_{s}$, or $\left\vert \vec{U}_{\mathrm{m}}^{(1)}\right\vert \approx \left\vert \Omega^{(1)}\right\vert R_{s}$, where $R_{s}$ is the radius of the star. We thus have, using the less favorable condition $\left\vert \vec{U}_{\mathrm{m}}^{(1)}\right\vert \approx\left\vert \Omega ^{(1)}\right\vert R_{s}$, that $f\approx\left\vert \vec{U}_{\mathrm{m}}^{(0)}\right\vert /\left( \Omega_{0}R_{s}\right)$.  

In this way, for $\left\vert \vec{U}_{\mathrm{m}}^{(0)}\right\vert \ll \Omega
_{0}R_{s}$, which is well satisfied in the case of the Sun, the model equation for the perturbed meridional mass flow is simplified to 
\begin{equation}
\rho \vec{U}_{\mathrm{m}}^{(1)}\cdot \nabla \left( \Omega
^{(0)} r^{2}\sin ^{2}\theta\right) =
\frac{1}{\mu _{0}}\left[ \nabla \times \left( A_{\phi}\,\textbf{e}
_{\phi }\right) \right] \cdot \nabla \left( r\sin \theta B_{\phi}\right) ,
\label{Um1a}
\end{equation}
which must be complemented with the meridional mass flow conservation
written in terms of a stream function $\psi $, 
\begin{equation}
\rho \vec{U}_{\mathrm{m}}^{(1)}=\frac{1}{r\sin \theta }\nabla \psi \times \textbf{e}_{\phi }.  \label{Um1b}
\end{equation}  

We note that $\vec{U}_{\mathrm{m}}^{(1)}$ affects directly $\vec{S}$, which has an explicit dependence on $\vec{U}$. The corresponding modification of $\alpha $ in Equation \ref{dAdt} requires further elaboration, which we consider next.

As mentioned above, according to the derivations in \cite{Sraibman} the $\alpha $ coefficient can be expressed in terms of the radial cylindrical component of the mean vorticity, $\omega _{s}$, 
\begin{equation}
\alpha =\frac{\lambda ^{2}\varkappa }{24s}\omega _{s},
\end{equation}
with $0<\varkappa <1$ an adjustable parameter of the model. 
The main, base contribution to $\omega_{s}$ comes from the differential rotation because, for the axisymmetric case considered,
\begin{equation}
\omega _{s}^{(0)}=-\frac{\partial U_{\phi }^{(0)}}{\partial z}=\sin
\theta \left[ \sin \theta \frac{\partial \Omega }{\partial \theta }-r\cos
\theta \frac{\partial \Omega }{\partial r}\right] .  \label{ws0}
\end{equation}
When one takes into account the modification to the base flow $\vec{U}_{\mathrm{m}}^{(1)}$, a new contribution $\omega _{s}^{(1)}$ is generated given place to a modified $\alpha $ coefficient. To evaluate this new contribution we start with the evolution equation of $\omega _{s}$, given by, in cylindrical coordinates,
\begin{equation}
\frac{\partial \omega _{s}}{\partial t}+U_{s}\left( \frac{\partial \omega
_{s}}{\partial s}+\frac{\omega _{s}}{s}\right) +U_{z}\frac{\partial \omega
_{s}}{\partial z} = 2\Omega _{0} \frac{\partial U_{s}}{\partial z}
-\omega _{s}\frac{\partial U_{z}}{\partial z}+D, \label{ws} 
\end{equation}
where $D$ represents the subscale term. We note that, due to
axisymmetry, there is no contribution of the pressure in \ref{ws}, even in
a generic baroclinic case. Here again the base flow satisfies the stationary
version of Equation \ref{ws}, so that, keeping only the terms of first order we
get
\begin{equation}
\frac{\partial \omega _{s}^{(1)}}{\partial t}+U_{s}^{(0)}\left( \frac{
\partial \omega _{s}^{(1)}}{\partial s}+\frac{\omega _{s}^{(1)}}{s}\right)
+U_{z}^{(0)}\frac{\partial \omega _{s}^{(1)}}{\partial z}+\omega _{s}^{(1)}
\frac{\partial U_{z}^{(0)}}{\partial z}=K,  \label{ws1}
\end{equation}
where
\begin{equation}
K=2\Omega _{0} \frac{\partial
U_{s}^{(1)}}{\partial z}-\omega _{s}^{(0)}\frac{\partial U_{z}^{(1)}}{
\partial z}
-U_{s}^{(1)}\left( \frac{\partial \omega _{s}^{(0)}}{\partial s}+\frac{
\omega _{s}^{(0)}}{s}\right) -U_{z}^{(1)}\frac{\partial \omega _{s}^{(0)}}{
\partial z}+D^{(1)}  ,\label{S0inter}
\end{equation}
which contains, apart from the first order term of the subscale stress
contribution, $D^{(1)}$, terms that are known, either from the base flow or
from the solution to Equations \ref{Um1a} and \ref{Um1b}. As with Equation \ref{Lz}, $D^{(1)}$ can be generically written as $D^{(1)}=\frac{{\lambda }^{2}}{24}\nabla F^{(0)}\cdot \nabla G^{(1)}$, where $F$ and $G$ correspond to generic mean field magnitudes of the indicated order. As before, since for a mean field magnitude $F$, ${\lambda}\left\vert\nabla F\right\vert \approx F$, the $D^{(1)}$ term is, due to the factor $1/24$, small in general compared with the other terms, it can be neglected to have 
\begin{equation}
 K=2\Omega _{0} \frac{\partial
U_{s}^{(1)}}{\partial z}-\omega _{s}^{(0)}\frac{\partial U_{z}^{(1)}}{
\partial z}
-U_{s}^{(1)}\left( \frac{\partial \omega _{s}^{(0)}}{\partial s}+\frac{
\omega _{s}^{(0)}}{s}\right) -U_{z}^{(1)}\frac{\partial \omega _{s}^{(0)}}{
\partial z}  .\label{S0final}
\end{equation}
The solution to Equation \ref{ws1} with source term given by Equation \ref{S0final} thus provides, through Equation \ref{alfa}, a contribution to the $\alpha $ coefficient given by 
\begin{equation}
\alpha ^{(1)}=\frac{\lambda ^{2}\varkappa }{24s}\omega _{s}^{(1)}.
\label{alpha1}
\end{equation}

\section{Numerical Simulation}
      \label{S-simulation}      

In order to test the original model in \cite{Sraibman} with the addition of Equations \ref{Um1a}, \ref{Um1b}, \ref{ws1}, \ref{S0final}, and \ref{alpha1}, we have applied it to a star of radius $R_{s}$\ with solar parameters. Equations \ref{dAdt} and \ref{dBdt} are solved in a full spherical shell, extending in radius from $r_{0}$, located below the bottom of the convection zone ($r_{0}=0.57R_{s}$), to the surface ($r=R_{s}$). It is assumed that the deep radiative interior is a perfect conductor, so that $r_{0}$ is chosen deeper than the lowest extent of the region where the dynamo action is taking place, so that the boundary conditions at $r_{0}$ are 
\begin{equation}
A_{\phi}=0,\;\; \partial (rB_{\phi})/\partial r=0. 
\end{equation}
At the surface we consider 
\begin{equation}
\nabla ^{2}A_{\phi}-A_{\phi}/\left( r^{2}\sin ^{2}\theta \right) =0,\;\; B_{\phi}=0.
\end{equation}

A differential rotation profile fitting the helioseismology data is
used \cite{Schou1998,Charbonneau1999}: 
\begin{equation}
\Omega \left( r,\theta \right) =\Omega _{0}+\frac{1}{2}\left[ 1+\mathrm{erf}\left( 2\frac{r-r_{t}}{d_{t}}\right) \right] \left[ \Omega _{\mathrm{SCZ}}\left( \theta\right) -\Omega _{0}\right] , \label{Omegahs}
\end{equation}
where $\Omega _{\mathrm{SCZ}}(\theta )=\Omega _{\mathrm{EQ}}+a_{2}\cos ^{2}(\theta )+a_{4}\cos
^{4}(\theta )$ is the surface latitudinal rotation. The value of the angular
velocity of the rigidly rotating core is $\Omega _{0}=2\pi \times 432.8$
nHz. The other parameters are set to $r_{t}=0.7R_{s}$, $d_{t}=0.05R_{s}$, $
\Omega _{\mathrm{EQ}}=2\pi \times 460.7$ nHz, $a_{2}=-62.69$ nHz, and $a_{4}=-67.13$
nHz.

Considering the still-debated characteristics of the meridional circulation \cite{Cameron2017}, including possible multicell structure \cite{Zhao2013}, and magnitude \cite{Schad2013,Jackiewicz2015}, we have chosen an order zero meridional circulation consisting in a single cell per hemisphere, not penetrating below the tachocline, written in conservative form as 
\begin{equation}
\rho \vec{U}_{\mathrm{m}}^{(0)}=\frac{1}{r\sin \theta }\nabla \Psi \times \textbf{e}_{\phi },
\end{equation}
with
\begin{equation}
\Psi =\sigma \sin ^{3}\theta \cos \theta (R_{s}-r)(r-r_{p})\exp [\lambda
(R_{s}-r)],  \label{psiMC}
\end{equation}
where $\sigma =-200.0$ kg m$^{-2}$s$^{-1}$, $r_{p}$ is the allowed depth of meridional circulation, set at $r_{p}=0.7R_{s}$, and $\lambda
=8.0/R_{s}$. The mass density $\rho $ is calculated from a stellar model for the Sun, and approximated in the convective region by 
\begin{equation}
\rho =C_{0}\left( 1-r/R_{s}\right) ^{2},  \label{rhoCZ}
\end{equation}
with $C_{0}=2.3\times 10^{3}$ kg m$^{-3}$. 
Expressions \ref{psiMC} and \ref{rhoCZ} result in a meridional velocity
field very similar in shape to those obtained in numerical simulations \cite{Kitchatinov2017}, and also similar to the simplest circulation pattern of those used in kinematic dynamo models \cite{Hazra2014,Hung2015,Rempel2006b}. The magnitude of the latitudinal velocity near the surface is about $12\; \textrm{m} \; \textrm{s}^{-1}$.

The small-scale magnetic diffusivity was modeled using Smagorinsky's
expression \cite{Scotti1993}, 
\begin{equation}
\eta (r,\theta )=C_{s}^{2}\Delta r\Delta \theta r\sqrt{2S_{kl}S_{kl}},
\end{equation}
where $S_{kl}$ is the large-scale rate of strain due to the meridional and
azimuthal large-scale velocity field, $\Delta r$ and $\Delta \theta $ the
local radial and latitudinal grid sizes, respectively. The value of the Smagorinsky's coefficient $C_{s}$ is usually taken to be between 0.1 and 0.2 \cite{Pope2000}. We have set it as $C_{s}=0.13$. 

Given the small magnitude of $\alpha ^{(1)}$ relative to that of $\alpha ^{(0)}$, the value of the parameter $\varkappa$ of $\alpha ^{(1)}$ was set to 1, its maximum possible value, leaving only the value of the parameter $\varkappa$ of $\alpha ^{(0)}$ as a free parameter, which was set to $\varkappa =0.26$ in order to obtain periods close to the observed 11-year (semi) period of the magnetic cycle (periods between about 8 and 14 years were obtained in the range of values of $\varkappa$ that resulted in cyclic dynamics). 

As initial condition a purely dipolar magnetic field, with about 10 G magnitude at the surface level, was used.

\begin{figure}[ht]
		\centerline{\includegraphics[bb=0 0 710 859,width=454pt]{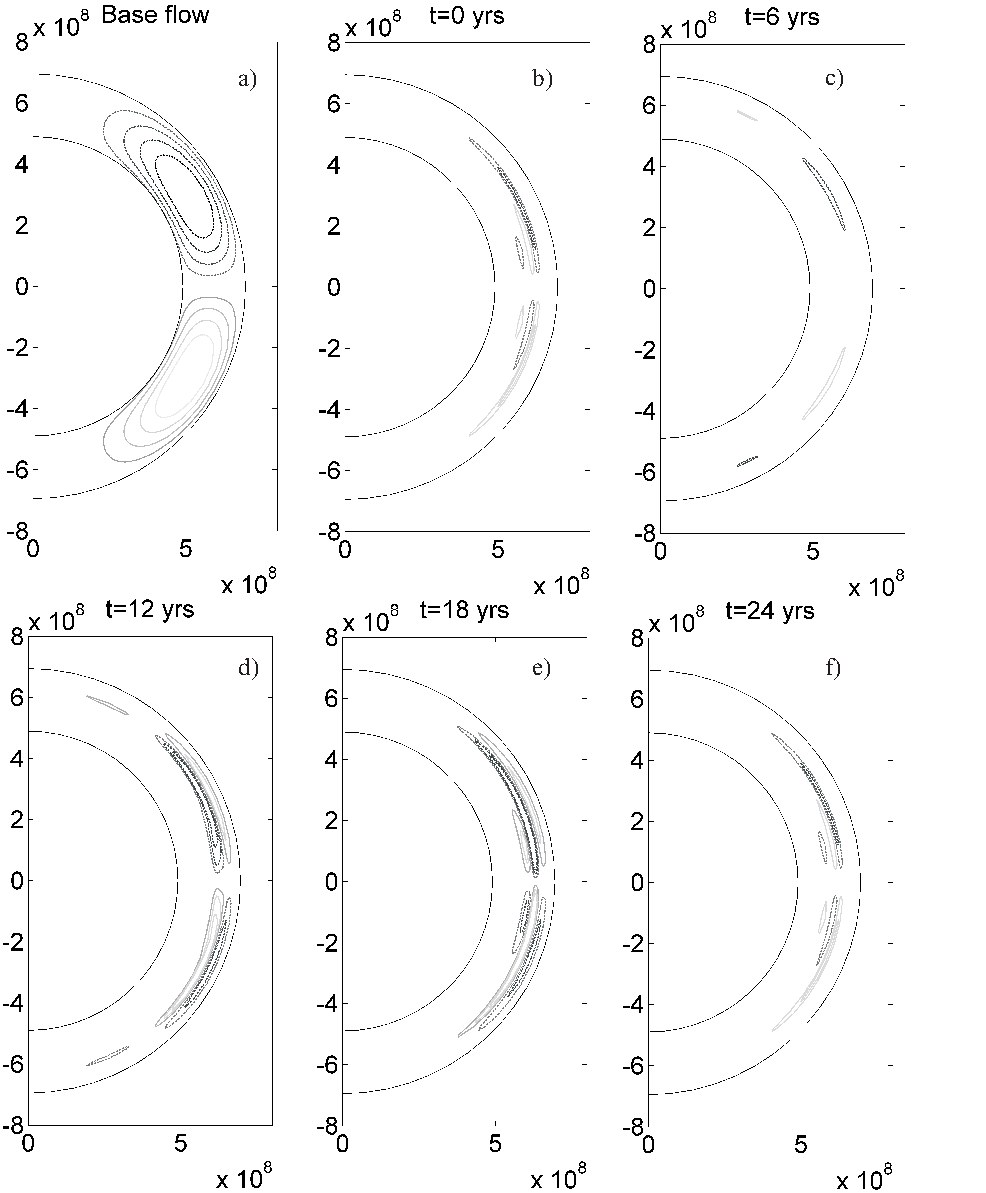}}
    \caption{Meridional circulation. Panel a): Mass-flow lines of the imposed base meridional flow. The increment between adjacent lines is $2\; \textrm{kg}\; \textrm{s}^{-1}$. Panels b) to f): Mass-flow lines of the meridional circulation induced by the Lorentz force at different times. The increment between adjacent lines is $0.02\; \textrm{kg}\; \textrm{s}^{-1}$.}
    \label{circulation}
\end{figure}

\begin{figure}[ht]
		\centerline{\includegraphics[bb=0 0 1113 1330,width=454pt]{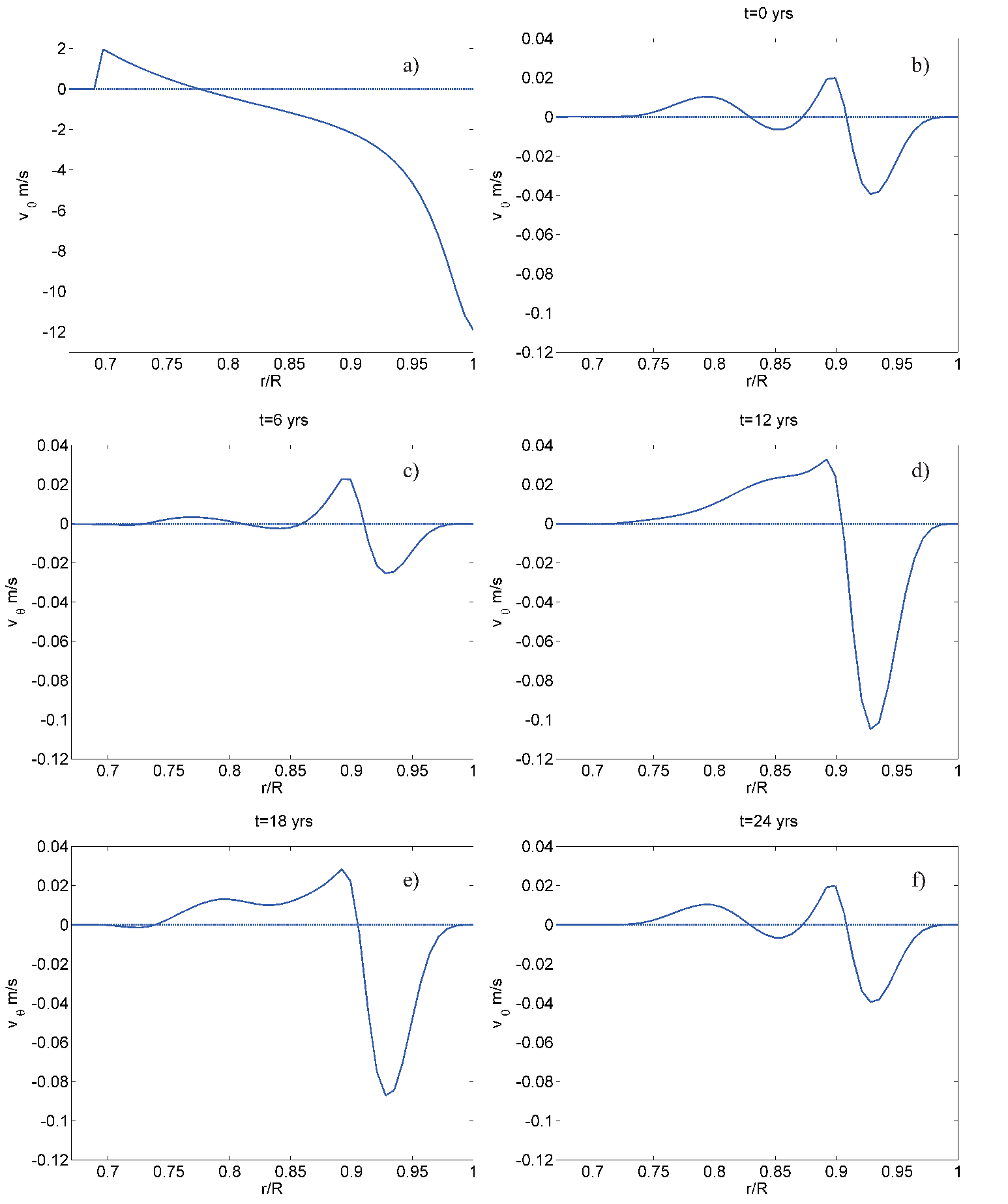}}
    \caption{Meridional velocity profiles at $45^{\circ}$ north latitude. Panel a): Corresponding to the base meridional flow. Panels b) to f): Corresponding to the meridional circulation induced by the Lorentz force at different times.}
    \label{circulation2}
\end{figure}

Figure~\ref{circulation} displays in the first panel the imposed meridional flow to be used in the whole simulation, and in the rest of the panels the induced circulation at different times. In this figure we indicate with 
$t=0$ an arbitrary time after a few hundred years of simulated time, where
one can see three main cells in low latitudes and one cell in middle to high
latitudes, with opposite signs in each hemisphere. In the following times it
can be seen that the pattern of the induced circulation is modified, both in
the number of cells and their position, but at year 24 the pattern is very close to that at $t=0$. We note that the highest magnitude of the induced poloidal velocity is slightly above $0.1\; \textrm{m}\; \textrm{s}^{-1}$, that occurs where the imposed velocity is close to $2\; \textrm{m}\; \textrm{s}^{-1}$, resulting in about 5\% variation.
The corresponding profiles of the meridional velocity at $45^{\circ}$ north latitude are displayed in Figure~\ref{circulation2}.  

\begin{figure}[ht]
		\centerline{\includegraphics[bb=0 0 372 563,width=454pt]{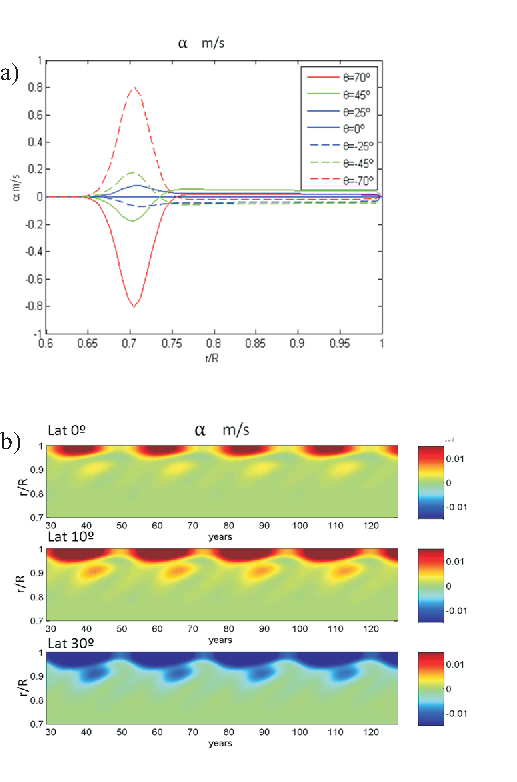}}
    \caption{$\alpha$ coefficients. a)  $\alpha ^{(0)}$ coefficient
for different latitudes. b) The induced $\alpha ^{(1)}$ as function of radius and time for three different latitudes.}
    \label{alphas}
\end{figure}

In Figure~\ref{alphas} the radial profile of the $\alpha ^{(0)}$ coefficient is shown for different latitudes in the first panel. These values are constant in time and correspond to Equation \ref{alfa} in which expressions \ref{ws0} and \ref{Omegahs} are used. In the rest of the panels we show the induced $\alpha^{(1)}$ given by Equations \ref{alpha1} and \ref{ws1} as function of radius and time for three different latitudes. As can be seen, $\alpha^{(1)}$ varies in time with a close to 24-year cycle near the surface, and with a close to 12-year cycle in the middle of the convection zone. The maximum value of $\alpha ^{(0)}$ is about $1\;  \textrm{m}\;  \textrm{s}^{-1}$, whereas the corresponding value of  $\alpha^{(1)}$ is about $1\;  \textrm{cm} \; \textrm{s}^{-1}$. 

\begin{figure}[ht]
		\centerline{\includegraphics[bb=0 0 576 433,width=454pt]{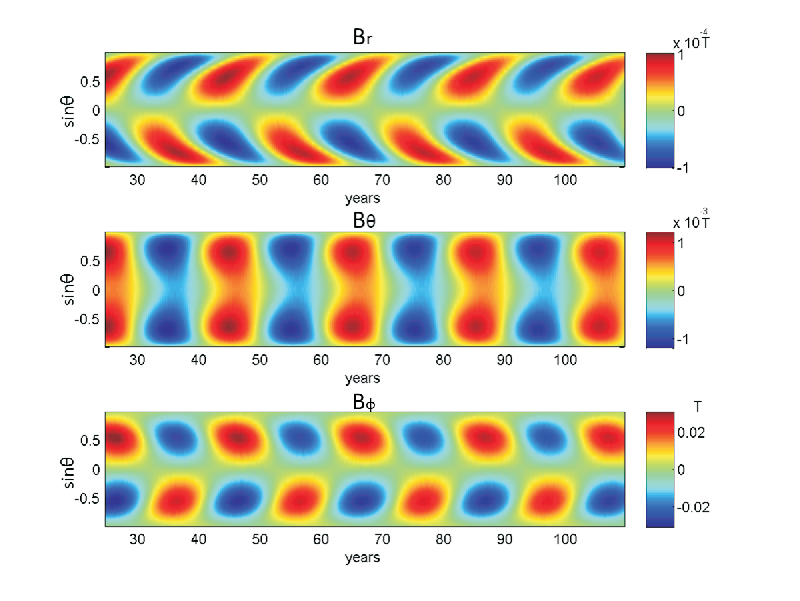}}
    \caption{Magnetic field components as functions of latitude and time. Top and middle panels: meridional components just below the surface. Bottom panel: azimuthal component just above the tachocline.}
    \label{fields}
\end{figure}

The evolution of the magnetic field components are presented in Figure~\ref{fields}. The top and middle panels show the poloidal components just below the surface as function of latitude and time. The bottom panel presents the toroidal component just above the tachocline as function of latitude and time. In this figure one can appreciate that the polarity reversals in all components of the magnetic field occur every twelve years, approximately, and that the amplitude of each component stays bounded at constant values. Amplitude values at the surface are about 1 G for $B_{r}$, 10 G for $B_{\theta}$, and 400 G for $B_{\phi}$. The amplitude of $B_{\phi}$ just above the tachocline is about 200 G. Also, the toroidal field shows a shift of its maximum magnitude from middle latitudes towards the equator, and the radial field a poleward migration at high latitudes. There is also a phase lag between $B_{r}$ and $B_{\phi}$ leading to a negative correlation between them.

\begin{figure}[ht]
		\centerline{\includegraphics[bb=0 0 1025 490,width=454pt]{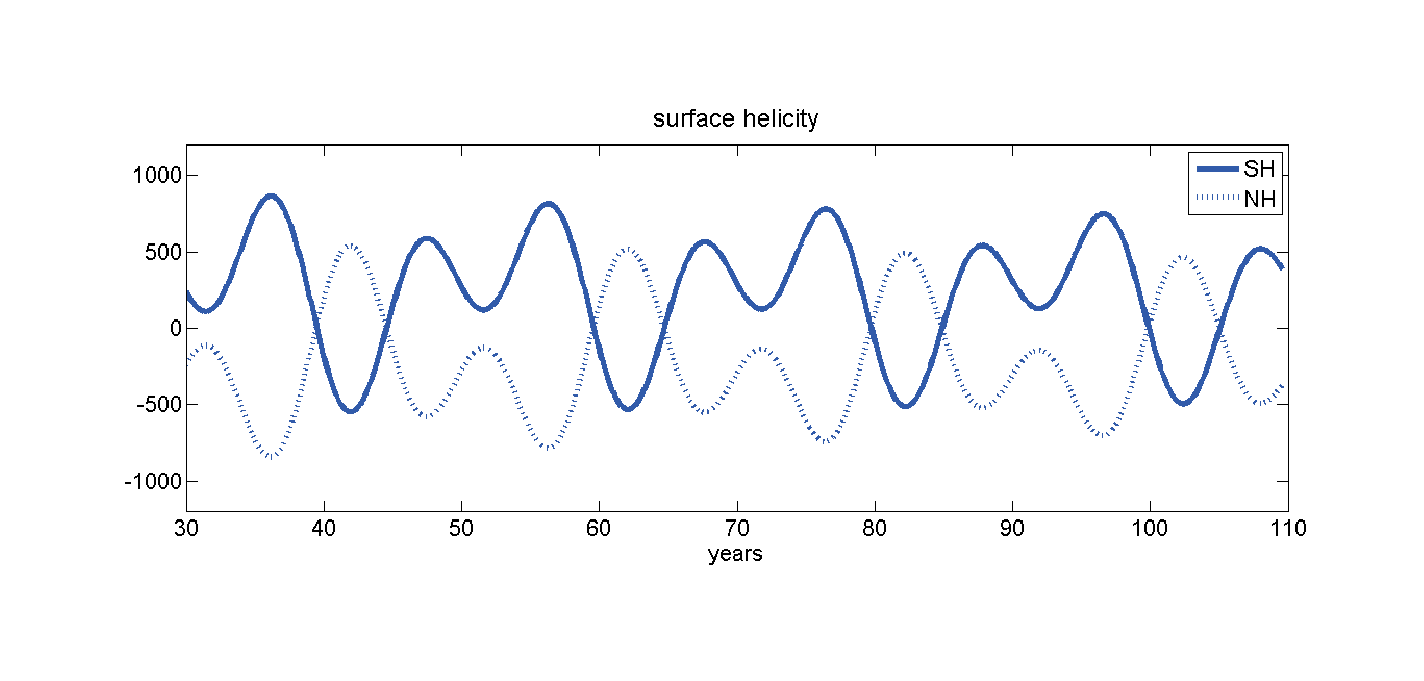}}
    \caption{Magnetic helicity density at the surface averaged over each hemisphere as a function of time.}
    \label{helicity}
\end{figure}

In Figure~\ref{helicity} the magnetic helicity density (per unit volume) on the surface layers of the star, averaged over each hemisphere, as function of time is displayed. Although the averaged helicity varies with time and even changes sign, the dominant sign is negative (positive) in the northern (southern) hemisphere. 

\begin{figure}[ht]
		\centerline{\includegraphics[bb=0 0 421 316,width=454pt]{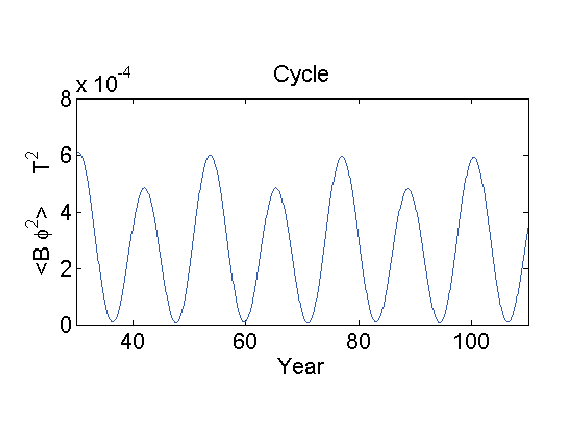}}
    \caption{Toroidal field just above the tachocline, squared and averaged over all latitudes as a proxy of cycle magnitude, as function of time.}
    \label{intensity}
\end{figure}

Finally, in Figure~\ref{intensity} we show, as a proxy of cycle magnitude, the toroidal field above the tachocline, squared and averaged over all latitudes as function of time. As a consequence of the induced meridional circulation having a 24-year period, the intensity of the 12-year cycle alternates in amplitude, a characteristic reminiscent of the Gnevyshev-Ohl Rule, or Even-Odd Effect \cite{Gnevyshev1948,Hathaway2015}. 

\begin{figure}[ht]
		\centerline{\includegraphics[bb=0 0 577 433,width=454pt]{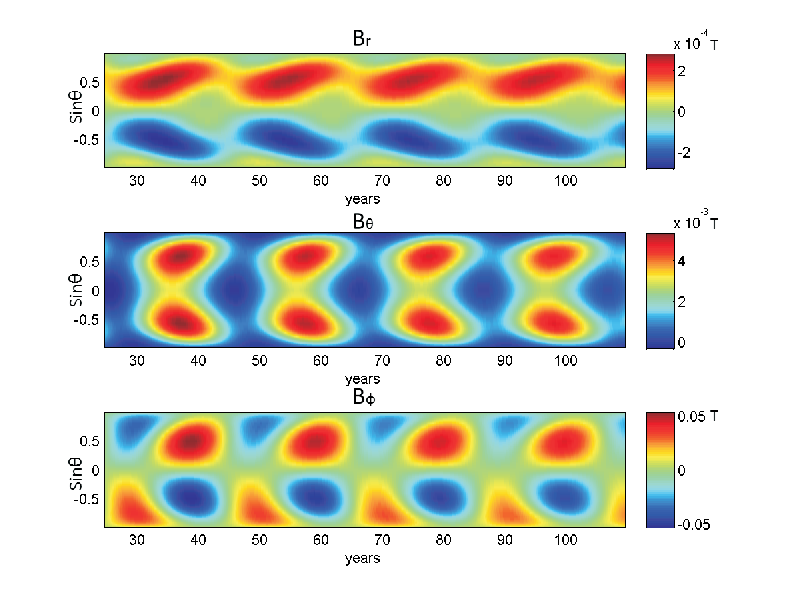}}
    \caption{Magnetic field components as functions of latitude and time for the simulation without feedback. Top and middle panels: meridional components just below the surface. Bottom panel: azimuthal component just above the tachocline.}
    \label{sinfeed}
\end{figure}

\begin{figure}[ht]
		\centerline{\includegraphics[bb=0 0 577 433,width=454pt]{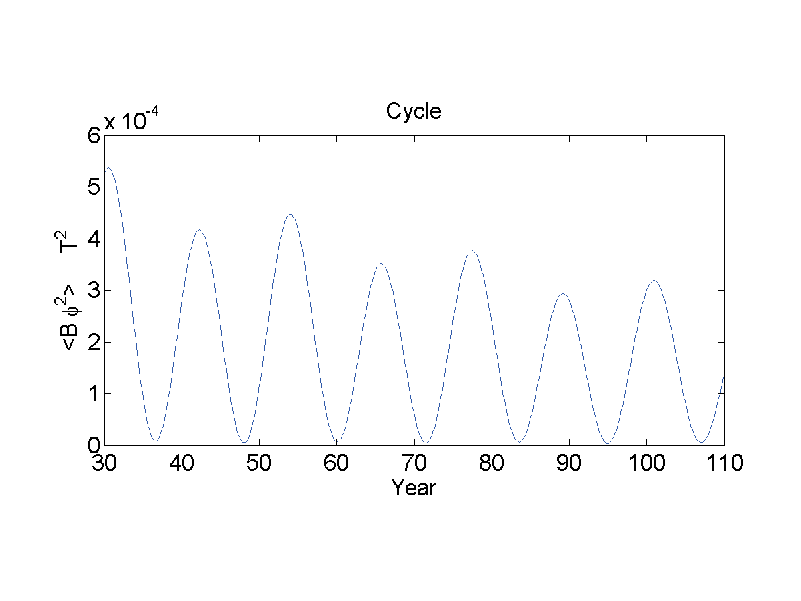}}
    \caption{Toroidal field just above the tachocline for the simulation without feedback, squared and averaged over all latitudes as a proxy of cycle magnitude, as function of time.}
    \label{cyclesinfeed}
\end{figure}

In order to test the effect of feedback, we have  run a simulation with exactly the same parameters and initial conditions, but without feedback effects. The magnetic field in this case, presented in Figure~\ref{sinfeed}, shows a decaying, cyclic dynamics in which the poloidal component has periodic variations of amplitude, without reversals, whereas the toroidal component reverses periodically, but with one polarity markedly stronger than the other.  
The decay of the magnetic field when no feedback is included can be better seen in Figure~\ref{cyclesinfeed}, in which the toroidal component, averaged over latitude, is shown as a function of time. 

\section{Conclusion} 
  \label{S-conclusion}

We have presented a model for an axisymmetric, large-scale dynamo that includes the meridional flow modification due to the back reaction of the magnetic field as a perturbation to imposed base flows driving the dynamo. The other condition required to close the equations is that the magnitude of the base meridional flow is small compared to the rotational velocity. Conditions that are well verified in the case of the Sun. These assumptions allow one to derive a closed model starting from the dynamic equations, using a previously presented formalism that requires parametrization only of scales much smaller than those resolved. In the presented simulation of a star with solar-like parameters the induced flow is about an order of magnitude smaller than the imposed flow, thus giving support to the formalism used. This relatively small modification of the base flow by the magnetic field suffices to limit and sustain the field amplitudes to realistic values, hundreds of gauss for the toroidal field and tens of gauss for the meridional component. The periodic dynamics also showing solar-like characteristics. The model with the same parameters and initial condition, but without feedback effects, shows a decaying magnetic field whose dynamics is different from that in the case with feedback, with a poloidal field of periodically varying magnitude, without reversals, and a weakly reversing toroidal component.  

\section*{Appendix: Expressions of $\vec{S}$ Terms}  

The explicit expression of the term $S_{\phi }$ that enters Equation \ref{dAdt} is 
\begin{eqnarray}
S_{\phi } =\frac{{\lambda }^{2}\,}{24\,r^{2}}\left[ -{B}_{{r}}{U}_{{
\theta }}-{U}_{{r}}\partial _{\theta }{B}_{{r}}-{U}_{{\theta }}\partial
_{\theta }{B}_{{\theta }}+{B}_{{r}}\partial _{\theta }{U}_{{r}}\right.\nonumber  \\ +\partial _{\theta }{B}_{{\theta }}\partial _{\theta }{U}_{{r}}-\partial
_{\theta }{B}_{{r}}\partial _{\theta }{U}_{{\theta }}+{B}_{{\theta }}\left( {
U}_{{r}}+\partial _{\theta }{U}_{\theta }\right) \  \nonumber \\
\left. +r^{2}\,\left( \partial _{r}{B}_{{\theta }}\partial _{r}{U}_{{r}
}-\,\partial _{r}{B}_{{r}}\partial _{r}{U}_{\theta }\right) \right] .\nonumber
\end{eqnarray}

On the other hand, it is convenient to express the terms depending on  $\vec{S}$ in Equation \ref{dBdt} by separating them in a part associated to the differential rotation and another one to the meridional flow. The part corresponding to the differential rotation is written as
\begin{eqnarray}
\left( \nabla \times \vec{S}\right) ^{\Omega }\cdot \textbf{e}
_{\phi } =\frac{\lambda ^{2}}{24r^{2}}\left[ F_{r}B_{r}+F_{\theta
}B_{\theta }+F_{r\theta }\partial _{r}B_{\theta }\right. \nonumber  \\
\left. +F_{\theta r}\partial _{\theta }B_{r}+F_{\theta \theta }\partial
_{\theta }B_{\theta }\right] , \nonumber
\end{eqnarray}
where
\begin{eqnarray}
F_{r}&=&3\cos \theta \partial _{\theta }\Omega +\sin \theta \left( \partial
_{\theta \theta }\Omega +r\partial _{r}\Omega -2r^{2}\partial _{rr}\Omega
\right) ,\nonumber \\
F_{\theta }&=&\frac{3+\cos 2\theta }{2}\csc \theta \partial _{\theta
}\Omega -r\sin \theta \partial _{r\theta }\Omega -r^{2}\cos \theta \partial
_{rr}\Omega ,\nonumber \\
F_{r\theta }&=&r^{2}\sin \theta \partial _{r\theta }\Omega , \nonumber\\
F_{\theta r}&=&r\cos \theta \partial _{r}\Omega -\sin \theta \partial
_{\theta }\Omega +r\sin \theta \partial _{r\theta }\Omega , \nonumber\\
F_{\theta \theta }&=&\cos \theta \partial _{\theta }\Omega +\sin \theta
\partial _{\theta \theta }\Omega -r^{2}\sin \theta \partial _{rr}\Omega .\nonumber
\end{eqnarray}
The part associated to the meridional flow is given by
\begin{eqnarray}
\left( \nabla \times \vec{S}\right) ^{U}\cdot \textbf{e}_{\phi }
=-\frac{\lambda ^{2}}{24r^{2}}\left[ G_{r}\partial _{r}B_{\phi}+G_{\theta
}\partial _{\theta }B_{\phi}+G_{r\theta }\partial _{r\theta }B_{\phi}\right.  \nonumber \\
\left. +G_{rr}\partial _{rr}B_{\phi}+G_{\theta \theta }\partial _{\theta \theta
}B_{\phi} \right] ,\nonumber 
\end{eqnarray}
where
\begin{eqnarray}
G_{r}&=&U_{r}+U_{\theta }\cot \theta +r\partial _{r}U_{r}+r^{2}\partial
_{rr}U_{r}+r\partial _{r\theta }U_{\theta },\nonumber \\
G_{\theta }&=&U_{r}\cot \theta +U_{\theta }\csc ^{2}\theta +\partial
_{\theta \theta }U_{\theta }-r\partial _{r}U_{\theta }+r\partial _{r\theta
}U_{r}, \nonumber\\
G_{r\theta }&=&-U_{\theta }+\partial _{\theta }U_{r}+r\partial
_{r}U_{\theta },\nonumber \\
G_{rr}&=&\partial _{r}U_{r}, \nonumber\\
G_{\theta \theta }&=&U_{r}\partial _{\theta }U_{\theta }.\nonumber
\end{eqnarray}

\subsection*{Acknowledgments}
We acknowledge the Consejo Nacional de Investigaciones Cient\'{i}ficas y T\'{e}cnicas (CONICET) and the University of Buenos Aires for institutional
support.
\bigskip

\bibliographystyle{iopart-num}
\bibliography{sraibman}  
\clearpage

\end{document}